\documentstyle[prl,aps,psfig,multicol]{revtex}
\begin{document}
\draft
\title{Effect of Tilted Magnetic Field on the Anomalous $H=0$ Conducting 
Phase\\ in High-Mobility Si MOSFETs}
\author{S.~V.~Kravchenko, D.~Simonian, and M.~P.~Sarachik}
\address{Physics Department, City College of the City University of New York, 
New York, New York 10031}
\author{A.~D.~Kent}
\address{Department of Physics, New York University, New York, New York 10003}
\author{V.~M.~Pudalov}
\address{Institute for High Pressure Physics, Troitsk, 142092 Moscow District, 
Russia}
\date{\today}
\maketitle
\begin{abstract}
The suppression by a magnetic field of the anomalous $H=0$ conducting phase in 
high-mobility silicon MOSFETs is independent of the angle between the field 
and the plane of the 2D electron system.  In the presence of a parallel field 
large enough to fully quench the anomalous conducting phase, the behavior is 
similar to that of disordered GaAs/AlGaAs heterostructures: the system 
is insulating in zero (perpendicular) field and exhibits reentrant 
insulator-quantum Hall effect-insulator transitions as a function of  
perpendicular field.  The results demonstrate that the suppression of the 
low-$T$ phase is related only to the electrons' spin.
\end{abstract}
\pacs{PACS numbers: 71.30.+h, 73.40.Qv, 73.40.Hm}
\begin{multicols}{2}

According to the one-parameter scaling theory of localization for 
non-interacting electrons\cite{gang}, a two-dimensional electron system (2DES) 
is always insulating at sufficiently large length scales ({\it i.e.}, in the 
limit of zero temperature) in the absence of a magnetic field.  In 
high-mobility silicon metal-oxide-semiconductor field-effect transistors 
(MOSFETs), however, a metal-insulator transition has been observed at a 
critical electron density, $n_c\sim10^{11}$~cm$^{-2}$, and a $H=0$ conducting 
phase has been shown to exist below 1~K\cite{krav}.  Similar critical behavior 
has been reported in a p-type SiGe quantum well\cite{col} and in the hole gas 
in GaAs/AlGaAs heterostructures\cite{danny,pepper}.  At low carrier densities, 
the interaction energy in these systems is more than an order of magnitude 
larger than the Fermi energy, so that one does not expect the 
non-interacting theory of localization\cite{gang} to be applicable in its 
simplest form.

In a disordered 2DES, Khmel'nitskii\cite{khmel} predicted that the extended 
states that exist at the centers of each Landau level in large perpendicular 
magnetic fields should ``float'' up in energy as $H_{\perp}\rightarrow0$, 
leading to an insulating phase at $H=0$.  Consistent 
with this expectation, insulating behavior has been observed in low-density, 
strongly disordered 2DES in GaAs/AlGaAs heterostructures\cite{jiang,reentr}.  
In contrast, the low-density 2DES in high-mobility Si MOSFETs exhibits quite 
different behavior.  As $H_{\perp}\rightarrow0$, the extended states shift 
upward from the centers of the Landau levels\cite{shashka}, as expected.  
However, instead of ``floating'' up indefinitely with decreasing magnetic 
field, the states apparently combine at the Fermi level\cite{shashka,pudalov}, 
giving rise to anomalous field dependence of $\rho_{xx}$ in small magnetic 
fields first reported in Ref.~\cite{diorio} and shown in the inset to 
Fig.~\ref{1}. This behavior is a puzzle, and its physical origin has 
remained unclear.

We have recently shown that the anomalous low-density/low-temperature 
conducting phase in silicon MOSFETs is suppressed by a magnetic field applied 
parallel to the 2D plane of the electrons\cite{para,pud}: as shown in 
Fig.~2 in Ref.\cite{para}, the resistivity increases by several orders of 
magnitude as the parallel magnetic field is increased to 
$H_{||}\sim20$~kOe, above which it saturates to a value that is approximately 
independent of magnetic field.  This prompted us to suggest that the 
enigmatic  behavior in small perpendicular fields is associated with the 
quenching of a low temperature conducting phase by a perpendicular field (see 
inset to Fig.~\ref{1}) just as it is quenched by a parallel field (see Fig.~2 
in Ref.\cite{para}).  We suggested further that the magnetic field suppression 
of the anomalous conducting phase in silicon MOSFETs is associated only with 
the electrons' spin, and is the same for any angle between the field and the 
2D plane of the electrons.

From measurements of the resistivity as a function of a magnetic field 
applied at different angles with respect to the plane of the electrons, we 
demonstrate in this Letter that:  (i)~A magnetic field suppresses 
the anomalous $H=0$ conducting phase in high-mobility silicon MOSFETs 
independently of the angle between the field and the plane of the electrons, 
thereby firmly establishing that the suppression of this phase is associated 
only with the electrons' spins.  (ii)~In the presence of a parallel field 
sufficiently large to quench the anomalous conducting phase in high-mobility 
silicon samples, the resistivity exhibits as a function of perpendicular field 
all the now-familiar features found in disordered, low-mobility GaAs/AlGaAs 
heterostructures\cite{jiang,reentr}: a giant negative magnetoresistance at low 
$H_{\perp}$, the quantum Hall effect (QHE) at Landau level filling factors 
$\nu=2$ and 1, and insulating behavior at higher $H_{\perp}$.  We also show 
that: (iii)~The suppression of the anomalous conducting phase is not 
associated with a simple change in mobility or electron density, both of which 
are essentially unaltered by the magnetic field; and (iv)~The multiple valleys 
that are
\vbox{
\vspace{0.4in}
\hbox{
\hspace{.4in}
\psfig{file=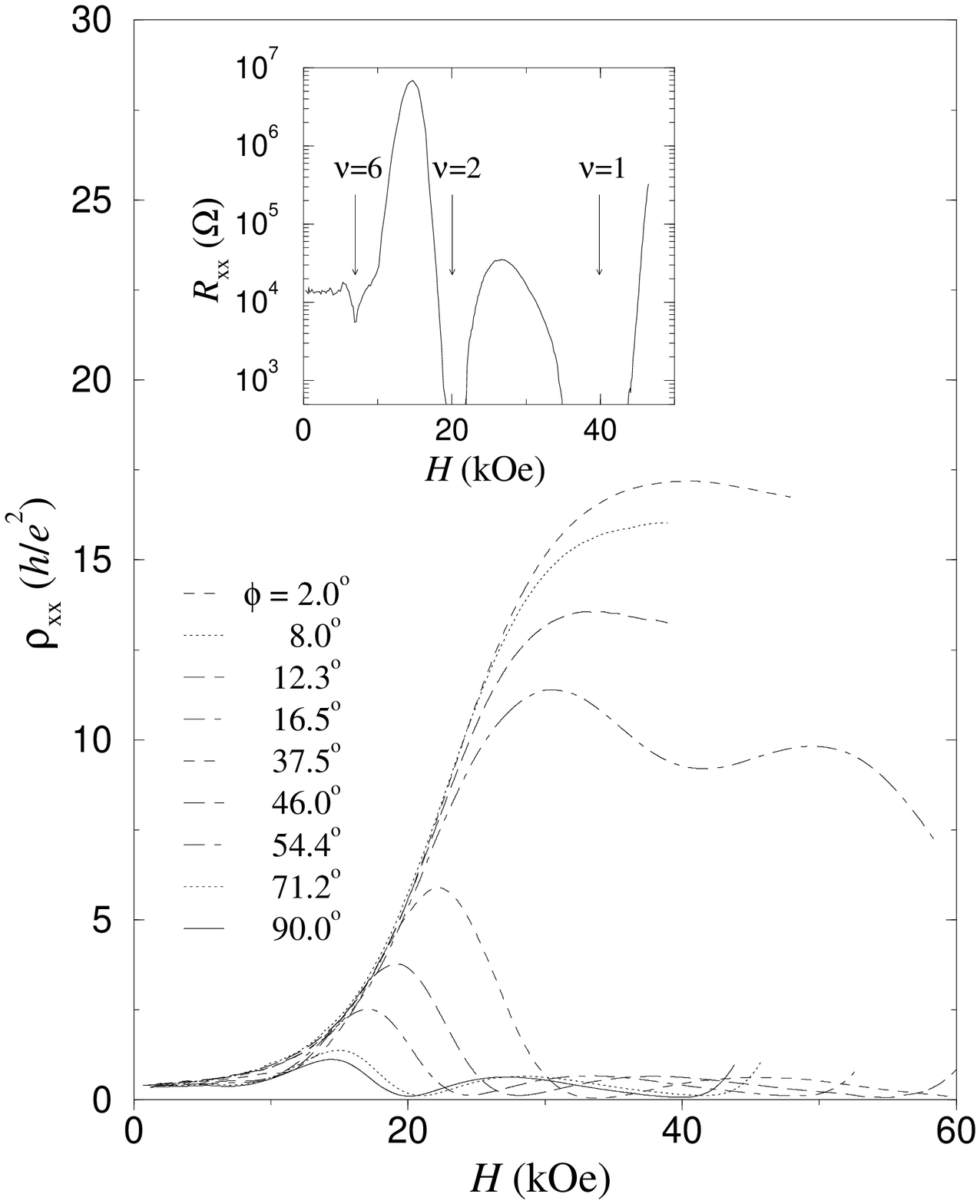,width=3.1in,bbllx=1.5in,bblly=1in,bburx=7.75in,bbury=9.25in,angle=0}
}
\vspace{0.15in}
\hbox{
\hspace{-0.15in}
\refstepcounter{figure}
\parbox[b]{3.4in}{\baselineskip=12pt \egtrm FIG.~\thefigure.
Resistivity as a function of the total magnetic field for a high-mobility
sample~B, at $T=0.36$~K and $n_s=1.0\times10^{11}$~cm$^{-2}$, for nine 
angles $\phi$ between the magnetic field and the inversion layer. $\rho_{xx}$
deviates from the ``main'' curve at smaller magnetic 
fields as $\phi$ is increased. (The inset shows the resistance of sample~A at 
$T=35$~mK as a function of a perpendicular magnetic field. At the electron 
density of $n_s=9.3\times10^{10}$~cm$^{-2}$ this sample is in the 
conducting state at $H=0$. QHE resistivity minima at filling 
factors $\nu=$1, 2, and 6 are shown by arrows. The resistivity 
decreases rapidly as $H\rightarrow0$, approaching a finite value
contrary to the expected insulating behavior).
\vspace{0.10in}
}
\label{1}
}
}
peculiar to the conduction band of silicon are not responsible for the 
low-temperature conducting phase, which is suppressed the same way by a field 
applied at any angle.

The three silicon MOSFET samples used for these studies have peak mobilities 
at 4.2 K of $\mu^{max}_{4.2K}\approx30,000$~cm$^2$/Vs (sample~A), 
$25,000$~cm$^2$/Vs (sample~B), and $8,000$~cm$^2$/Vs (sample~C).  
Four-terminal DC transport measurements were taken as a function of a magnetic 
field applied at different angles with respect to the plane of the electrons.  
Two Si MOSFET samples were measured in a pumped $^3$He system equipped with a 
12-Tesla magnet and a manual sample rotator.  Sample~A was studied in a 
dilution refrigerator in a magnetic field oriented perpendicular to the 2D 
plane.  Excitation currents were between 0.01~nA and 10~nA; care was taken to 
ensure measurements were in the linear $I-V$ regime.

For a gate voltage that placed sample~B in the conducting state at $H=0$ with 
a resistivity of $\approx10$~k$\Omega$ at 360~mK, Fig.~\ref{1} shows the 
diagonal resistivity, $\rho_{xx}$, as a function of a magnetic field applied 
at different angles with respect to the plane of the 2DES.  For all angles, 
$\rho_{xx}(H)$ follows approximately the same curve up to some value of 
magnetic field, above which orbital effects leading to QH oscillations become 
dominant.  The resistivity deviates from the ``main'' curve at smaller magnetic 
fields as the angle between the field and the plane is increased:  the larger 
perpendicular component causes stronger orbital effects which become dominant 
at a lower total field.  We note that small differences in $\rho_{xx}(H)$ at 
$H\sim10$~kOe are associated with the emergence of a QHE minimum at filling 
factor $\nu=6$\cite{pudalov}, which deepens as the perpendicular component of 
the field gets larger.  The important feature is that the magnetoresistance 
is the same at all angles up to some field above which it is overwhelmed by 
orbital effects.  The anomalous $H=0$ conducting phase is thus suppressed 
in the same manner by a magnetic field applied at any angle.

This provides evidence that the conduction band valleys in 
silicon do not play an important role.  It has been shown\cite{valleys} 
that a field applied parallel to the plane of the 2DES in silicon MOSFETs does 
not affect the splitting of two conduction band valleys.  This splitting is 
enhanced in a perpendicular field due to exchange interactions, and is 
therefore expected to be a function of field orientation.  The absence of 
any angular dependence implies that valley-splitting is not responsible for 
the suppression of the low-temperature conducting phase by a magnetic field.  
We thus arrive at the important conclusion that it is the electrons' spin that 
plays a crucial role.  
Indeed, among the theoretical suggestions that have been offered as possible 
explanations of the conducting phase at $H=0$ 
\cite{fink,dobro,belki,philli,spinorb,bulutay,rice}, several involve electron 
spins\cite{fink,belki,philli,spinorb}.

We now verify explicitly that a magnetic field does not drive the 
sample into the insulating phase by simply reducing the electron 
mobility\cite{dirty}, or by reducing the electron density below its critical 
value.  Fig.~\ref{2} shows $\mu_{\text{4.2~K}}$ of high-mobility sample~B as a 
function of electron density in $H=0$ and in the presence of a parallel 
magnetic field, $H_{||}$=30~kOe.  These data establish that the mobility is 
essentially unaltered by a magnetic field.

The inset to Fig.~\ref{2} shows the resistance as a function of the 
perpendicular component of the magnetic field,
\vbox{
\vspace{0.15in}
\hbox{
\hspace{.4in}
\psfig{file=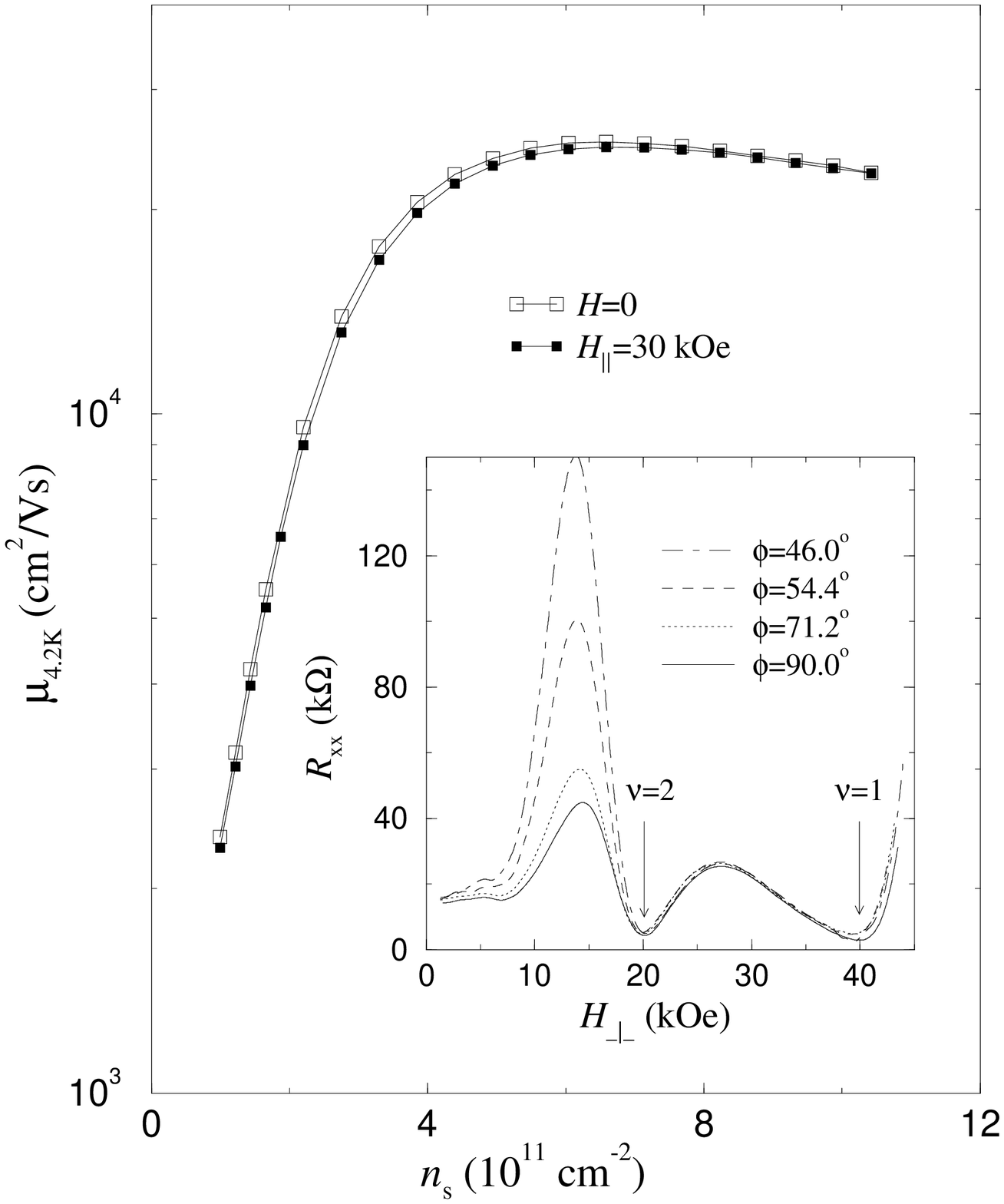,width=3.1in,bbllx=1.5in,bblly=1in,bburx=7.75in,bbury=9.25in,angle=0}
}
\vspace{0.15in}
\hbox{
\hspace{-0.15in}
\refstepcounter{figure}
\parbox[b]{3.4in}{\baselineskip=12pt \egtrm FIG.~\thefigure.
Mobility {\it vs} electron density for sample~B at $T=4.2$~K in zero magnetic 
field (open 
symbols) and $H_{||}=30$~kOe (closed symbols).  The inset 
shows $R_{xx}$ as a function of $H_{\perp}$ for four angles between the field and 2D plane; $T=0.36$~K and $n_s=1.0\times10^{11}$~cm$^{-2}$.
\vspace{0.10in}
}
\label{2}
}
}
$H_{\perp}=H$sin$\phi$, as the 
total field $H$ is swept at four different fixed angles with respect 
to the electron plane. Note that the parallel field, $H_{||}=H$cos$\phi$, 
varies along each curve and is different for different angles $\phi$.  The 
QHE minima occur at the same $H_{\perp}$ for all angles, corresponding to 
different values of the total field. This observation establishes that the 
magnetic field does not change the electron density in the inversion layer.  
The dramatic growth with angle of the $\rho_{xx}$ maximum at 
$H_{\perp}\sim15$~kOe can be understood by noting that the $H=0$ conducting 
state is quenched independently of the field orientation: at a fixed 
$H_{\perp}\approx15$~kOe, the total field increases with decreasing 
$\phi$, $H=H_{\perp}$(sin$\phi$)$^{-1}$, driving the sample closer to the 
insulating state.  Note that an anomalous growth with $H_{||}$ of 
the resistance peak between $\nu=1$ and $\nu=3$ has been observed in 
a p-Si/SiGe heterostructure\cite{dorozhka} and was attributed by the 
authors to the dependence of the ``insulating state width on the ratio between 
spin and cyclotron splittings''. We remark that the observation of a $H=0$ 
conducting state similar to that in high-mobility Si MOSFETs in this 
system\cite{col} suggests that the strong enhancement of the resistivity in 
p-Si/SiGe\cite{dorozhka} may instead be due to the magnetic field suppression 
of the anomalous conducting state in the same way as in Si MOSFETs.
\vbox{
\vspace{0.15in}
\hbox{
\hspace{.4in}
\psfig{file=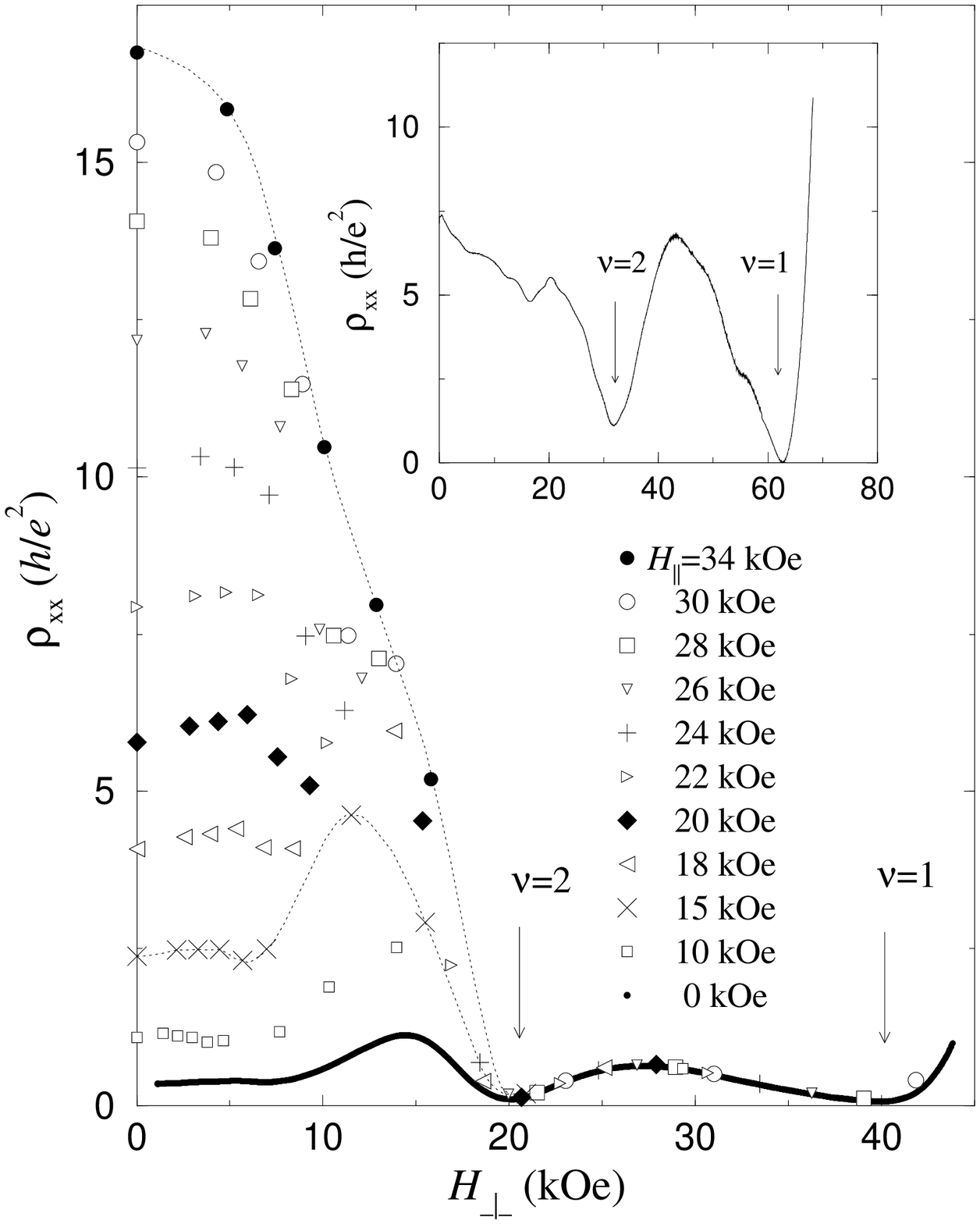,width=3.1in,bbllx=1.5in,bblly=1in,bburx=7.75in,bbury=9.25in,angle=0}
}
\vspace{0.15in}
\hbox{
\hspace{-0.15in}
\refstepcounter{figure}
\parbox[b]{3.4in}{\baselineskip=12pt \egtrm FIG.~\thefigure.
$\rho_{xx}$ of sample~B as a function of $H_{\perp}$ for different 
values of the parallel magnetic field; $T=0.36$~K and 
$n_s=1.0\times10^{11}$~cm$^{-2}$. The inset shows $\rho_{xx}(H_{\perp})$ 
for a low-mobility sample~C; $T=0.36$~K and $n_s=2.1\times10^{11}$~cm$^{-2}$.
\vspace{0.10in}
}
\label{3}
}
}

The diagonal resistivity, $\rho_{xx}$, is plotted as a function of 
$H_{\perp}$ in several fixed parallel magnetic fields in Fig.~\ref{3}. The 
lowest curve corresponds to $H_{||}=0$ and exhibits the anomalous behavior of 
high-mobility Si MOSFETs\cite{diorio}.  The peak at $H_{\perp}\approx15$~kOe 
is considerably smaller than that shown in the inset to Fig.~\ref{1} because 
of the higher measuring temperature (360~mK {\it vs} 35~mK).  The 
highest curve is the magnetoresistance of the sample in the insulating 
state, obtained by quenching the $H=0$ conducting state with a parallel field 
of 34~kOe.  In the ``quenched'' phase, high-mobility Si 
MOSFETs display the familiar reentrant behavior found 
in disordered, weakly interacting GaAs/AlGaAs heterostructures (see, 
{\it e.g.}, Fig.~2 in Ref.\cite{jiang}): the system has an initial large 
negative magnetoresistance, exhibits the quantum Hall effect at $\nu=2$ and 1, 
and becomes again insulating at $H\gtrsim42$~kOe. (However, the 
initial decrease in resistivity is considerably less sharp than in 
disordered GaAs/AlGaAs.) The gradual disappearance 
of the anomalous conducting phase is apparent at intermediate fields. Note 
that above $H_{\perp}\sim20$~kOe all the data collapse onto a single curve.  
This confirms once again that the anomalous phase is quenched by a magnetic 
field applied in any direction (including perpendicular).

The inset to Fig.~\ref{3} shows the diagonal magnetoresistivity 
$\rho_{xx}$ of the relatively low-mobility sample~C in a perpendicular field.  
No $H=0$ conducting phase was found in this sample. It is strongly 
insulating at $H_{\perp}=0$, and there is an appreciable negative 
magnetoresistance for $H_{\perp}\lesssim30$~kOe.  The $\nu=1$ and 
$\nu=2$ QHE minima in $\rho_{xx}$ are evident, followed at higher field by a 
transition to an insulator due to the crossing of the last extended state 
through the Fermi level at $H_{\perp}\gtrsim65$~kOe.  It is interesting that 
$\rho_{xx}$ {\it vs} $H_{\perp}$ for sample~C is qualitatively similar to the 
behavior of the high-mobility sample~B in a partially quenched phase.  
Moreover, sample~C exhibits a strong positive magnetoresistance as a function 
of $H_{||}$ (not shown).  This suggests that the anomalous low-temperature 
phase that is so evident in high-mobility samples is also present in a 
modified, partially quenched form in low-mobility, disordered samples.

In conclusion, we have shown that the suppression by a magnetic field of 
the $H=0$ conducting phase in high-mobility Si MOSFETs does not depend 
on the angle the field makes with the 2D electron plane. This provides strong 
evidence that valley splitting does not play an important role, and that the 
quenching of the anomalous conducting phase in two dimensions is associated 
with the electrons' spin.  We have also demonstrated explicitly that the 
suppression of the conductivity is not associated with a simple change 
in sample mobility or electron density, both of which are essentially 
unaffected by magnetic field.  In the presence of large parallel field, 
the ``quenched'' phase in high-mobility silicon MOSFETs exhibits the 
reentrant behavior of disordered, weakly interacting GaAs/AlGaAs 
heterostructures:  a large negative magnetoresistance and reentrant 
insulator-QHE-insulator transitions\cite{jiang,reentr}.

We are indebted to Robert Wheeler for generously supplying MOSFET 
devices fabricated in his laboratory.  We thank Veronika~Simonian for 
help in analyzing data using Mathematica$^{TM}$, Mark~Ofitserov 
for assistance with the experimental equipment, and Dan~Shahar for useful 
discussions.  This work  was supported by the US Department of Energy under 
Grant No.~DE-FG02-84ER45153.  A.\ K.\ was supported by NYU and an NYU Research 
Challenge Fund Grant. V.\ P.\ was supported by RFBR (grant 97-02-17387) and by 
INTAS.

\end{multicols}

\begin{references}
\bibitem{gang} E.\ Abrahams, P.\ W.\ Anderson, D.\ C.\ Licciardello, and 
T.\ V.\ Ramakrishnan, Phys.\ Rev.\ Lett.\ {\bf 42}, 673 (1979).
\bibitem{krav} S.\ V.\ Kravchenko, G.\ V.\ Kravchenko, J.\ E.\ Furneaux, 
V.\ M.\ Pudalov, and M.\ D'Iorio, Phys.\ Rev.\ B {\bf 50}, 8039 (1994); 
S.\ V.\ Kravchenko, W.\ E.\ Mason, G.\ E.\ Bowker, J.\ E.\ Furneaux, V.\ M.\ 
Pudalov, and M.\ D'Iorio, Phys.\ Rev.\ B {\bf 51}, 7038 (1995); 
S.\ V.\ Kravchenko, D.\ Simonian, M.\ P.\ Sarachik, W.\ Mason, and J.\ E.\ 
Furneaux, Phys.\ Rev.\ Lett.\ {\bf77}, 4938 (1996).
\bibitem{col} P.\ T.\ Coleridge, R.\ L.\ Williams, Y.\ Feng, and P.\ 
Zawadzki, to be published in Phys.\ Rev.\ B, Rapid Communications; see also 
preprint cond-mat/9708118.
\bibitem{danny} Y.\ Hanein, U.\ Meirav, D.\ Shahar, C.\ C.\ Li. D.\ C.\ Tsui, 
and H.\ Shtrikman, cond-mat/9709184.
\bibitem{pepper} M.\ Y.\ Simmons, A.\ R.\ Hamilton, M.\ Pepper, E.\ H.\ 
Linfield, P.\ D.\ Rose, and D.\ A.\ Ritchie, cond-mat/9709240.
\bibitem{khmel} D.\ E.\ Khmel'nitskii, Pis'ma Zh.\ Eksp.\ Teor.\ Fiz.\ {\bf 
38}, 454 (1983) [JETP Lett.\ {\bf 38}, 552 (1983)]; Phys.\ Lett.\ A {\bf 106}, 
182 (1984); see also R.\ B.\ Laughlin, Phys.\ Rev.\ Lett.\ {\bf 52}, 2304 
(1984); S.\ Kivelson, D.-H.\ Lee, and S.\ C.\ Zhang, Phys.\ Rev.\ B {\bf 46}, 
2223 (1992).
\bibitem{jiang} H.\ W.\ Jiang, C.\ E.\ Johnson, K.\ L.\ Wang, and S.\ T.\ 
Hannahs, Phys.\ Rev.\ Lett.\ {\bf 71}, 1439 (1993).
\bibitem{reentr} T.\ Wang {\it et al}, Phys.\ Rev.\ Lett.\ {\bf 72}, 709 
(1994); R.\ J.\ F.\ Hughes {\it et al}, J.\ Phys.\ Condens.\ Matter {\bf 6}, 
4763 (1994); D.\ Shahar, D.\ C.\ Tsui, and J.\ E.\ Cunningham, Phys.\ Rev.\ B 
{\bf 52}, R14~372 (1995); I.\ Glozman, C.\ E.\ Johnson, and H.\ W.\ Jiang, 
Phys.\ Rev.\ Lett.\ {\bf 74}, 594 (1995).
\bibitem{shashka} A.\ A.\ Shashkin, G.\ V.\ Kravchenko, and V.\ T.\ 
Dolgopolov, Pis'ma Zh.\ Eksp.\ Teor.\ Fiz.\ {\bf 58}, 215 (1993) [JETP Lett.\ 
{\bf 58}, 220 (1993)].
\bibitem{pudalov} V.\ M.\ Pudalov, M.\ D'Iorio, and J.\ W.\ Campbell, Surf.\ 
Sci.\ {\bf 305}, 107 (1994).
\bibitem{diorio} M.\ D'Iorio, V.\ M.\ Pudalov, and S.\ G.\ Semenchinsky, 
Phys.\ Lett.\ A {\bf 150}, 422 (1990).
\bibitem{para} D.\ Simonian, S.\ V.\ Kravchenko, M.\ P.\ Sarachik, and 
V.\ M.\ Pudalov, Phys.\ Rev.\ Lett.\ {\bf 79}, 2304 (1997).
\bibitem{pud} V.\ M.\ Pudalov, G.\ Brunthaler, A.\ Prinz, and G.\ Bauer, Pis'ma 
Zh.\ Eksp.\ Teor.\ Fiz.\ {\bf 65}, 887 (1997) [JETP Lett. {\bf 65}, 932 (1997)].
\bibitem{valleys} R.\ J.\ Nicholas, K.\ von\ Klitzing, and Th.\ Englert, 
Solid State Commun.\ {\bf 34}, 51 (1980), see also T.\ Ando, A.\ B.\ Fowler, 
and F.\ Stern, Rev.\ Mod.\ Phys.\ {\bf 54}, 437 (1982) and ref's therein.
\bibitem{fink} A.\ M.\ Finkel'shtein, Zh.\ Eksp.\ Teor.\ Fiz.\ {\bf 84}, 168 
(1983) [Sov. Phys. JETP {\bf57}, 97 (1983)].
\bibitem{dobro} V.\ Dobrosavljevi\'{c}. E.\ Abrahams, E.\ Miranda, and S.\ 
Chakravarty, Phys.\ Rev.\ Lett.\ {\bf 79}, 455 (1997).
\bibitem{belki} D.\ Belitz and T.\ R.\ Kirkpatrick, cond-mat/9705023.
\bibitem{philli} P.\ Phillips, Y.\ Wan, I.\ Martin, S.\ Knysh, and 
D.\ Dalidovich, submitted to Nature (London); cond-mat/9709168.
\bibitem{spinorb} V.\ M.\ Pudalov, cond-mat/9707076.
\bibitem{bulutay} C.\ Bulutay and M.\ Tomak, cond-mat/9707339.
\bibitem{rice} F.-C. Zhang and T.\ M.\ Rice, cond-mat/9708050.
\bibitem{dirty} Disappearance of the metallic phase with decreasing 
mobility has been reported in the first paper in Ref.\cite{krav} and demonstrated on a single sample with variable mobility in Ref.\cite{dragana}.
\bibitem{dragana} D.\ Popovi\'{c}, A.\ B.\ Fowler, and S.\ Washburn, Phys.\ 
Rev.\ Lett.\ {\bf 79}, 1543 (1997).
\bibitem{dorozhka} S.\ I.\ Dorozhkin, C.\ J.\ Emeleus, T.\ E.\ Whall, and 
G.\ Landwehr, Phys.\ Rev.\ B {\bf 52}, R11~638 (1995).
\end{references}
\end{document}